\font\amsfont=msxm10
\def\dal{\hbox{\amsfont\char'3}}  
\def\Real{{\mathchoice  
          {\textstyle \rm\hskip 0.2em R\kern -0.95em I\kern 0.55em}
          {\textstyle \rm\hskip 0.2em R\kern -0.95em I\kern 0.55em}
          {\scriptstyle \rm\hskip 0.15em R\kern -0.7em I\kern 0.3em}
          {\scriptstyle \rm\hskip 0.15em R\kern -0.7em I\kern 0.3em}
	  }}

\input psfig
\parindent=0pt
\parskip=\medskipamount

\centerline{\bf Generalised hyperbolicity in singular space-times}
\medskip
\centerline{C J S Clarke}
\centerline{Faculty of Mathematical Studies}
\centerline{University of Southampton}
\centerline{Southampton, SO17 1BJ, UK}
\bigskip
{\it Abstract}
A new concept analogous to global hyperbolicity is introduced, based
on test fields. It is shown that the space-times termed here
``curve integrable'' are globally hyperbolic in this new sense, and a
plausibility argument is given suggesting that the result applies to
shell crossing singularities. If the assumptions behind this last
argument are valid, this provides an alternative route to the
assertion that such singularities do not violate cosmic censorship.

\bigskip
{\bf 1. Introduction}

Penrose's strong cosmic censorship hypothesis [7] postulates that,
subject to genericness conditions, space-time will be globally
hyperbolic: i.e.\ strong causality holds and $J^{+}(p)\cap J^{-}(q)$ is 
compact for all $p, q$.
The context in which this is usually discussed [6] is that of
space-times where the metric is $C^{2-}$ (ensuring the existence of
unique geodesics in the classical sense). As result, a Lorentz
manifold with a metric that fails to be $C^{2-}$ at one point $p$ has
to be viewed as a space-time from which $p$ is to be deleted, which
usually results in a failure of global hyperbolicity and thus a
breakdown in cosmic censorship.

On the other hand, there are increasingly many examples emerging of
space-times that violate cosmic censorship in this way, but where
there is a well posed initial value problem for test fields. Global
hyperbolicity is sufficient, but not necessary for this. The
physically meaningful condition is not global hyperbolicity, but the
well-posedness of the field equations. This suggests that we need to
redefine the notion of hyperbolicity (or equivalently, of cosmic
censorship) so as to make direct reference to test fields, in
situations where a putative singularity $p$ is an internal point of a
Lorentz manifold $(M,g)$ with $g$ not being $C^{2-}$ at $p$. (A 
general
discussion of this idea is given in [2].)

We shall suppose throughout that we are working in a region in which there
exist local coordinates
(we regard $M\subset\Real^4$) in which $g_{ij}$ and $g^{ij}$ 
are
bounded but not necessarily continuous. This is not essential for many
of the points discussed below, but it is not my aim here to discuss
global issues. The definitions that follow below are
then made with respect to a choice of some particularl type of
physical field $\phi$ (e.g.\ massless scalar) satisfying $L\phi = 0$
for a second order differential operator $L= L(g)$.

There is an underlying conflict (which will not be resolved here) between the
view of geometrical general relativity, in which one deals
with differentiablity in the four-dimensional manifold, and the $3+1$
setting appropriate to the analysis of hyperbolic equations, in which
function-spaces are defined on three-dimensional hypersurfaces. The ideal
would be an integrated approach in which the full coupled Einstein-matter
equations were handled in a manner consistent with their hyperbolic
nature. In this paper I am the inhabiting a helf-way house, in which the metric
is being is being handled in terms of 4-D differentiability, while the
matter is being regarded as a test field used to probe the metric and is
described in $3+1$ terms. Thus for the case being considered where $L$ is second order, we shall take a
foliation of the region which establishes a particular diffeomorphism with
$\Real^3\times\Real$ and regard the
field $\phi$ as a map $\Phi:t\mapsto (\phi(.,t), \dot\phi(.,t))$ taking values
in an appropriate function space (defined by the norms in the next section)
on $\Real^3$, and $L$ will have the form $d\Phi/dt + A\Phi$ for a
three-dimensional differential operator $A$. For simplicity of notation,
however, I shall usually not distinguish explicitly between $\Phi$ and $\phi$.

The extension of the ``usual'' definition of an operator to the case
of a non-smooth metric is not always unambiguous, or even possible. In the case, however, of the
wave operator which will be treated here we can regard $\Phi(t)$ as
lying in $H^1(\Real^3)\times H^0(\Real3)$ ($H^i$ being the Hilbert
space of $i$ times differentiable functions. For smooth $g$ and for
$\psi,\phi\in C_0^\infty(\Real^4)\times C_0^\infty(\Real^4)$ we have that
$$
\int \psi (\dal \phi) d^4x = -\int\!\!\int\left[\dot\psi\left(g^{00}\dot\phi
+g^{0\alpha}\phi_{,\alpha}\right) + \psi_{,\beta}\left(g^{\beta0}
\dot\phi +
g^{\beta\alpha}\phi_{,\alpha}\right)\right]\sqrt{-g}d^3x\,dt.
$$
The $\Real^3$ integral on the right hand side defines, for fixed $\phi$, a linear
operator on $\psi$ which is bounded on $H^1(\Real^3)\times
H^0(\Real3)$, and is well defined for a general bounded invertible
$g$. Hence there is an element $A(\Phi(t))$ of this space
such that
$$
\int \psi (\dal \phi) d^4x = -\int \langle \Psi(t), A(\Phi(t))dt
$$
thus defining A (an unbounded operator on a dense domain) for general $g$.

In this context, I shall call $\phi$ a ``solution'' to $L\phi=0$ if there
exists a foliation with respect to which $d\Phi/dt + A\Phi=0$. Note that
this therefore does not imply that $\phi$ is $C^2$. A solution in this
sense is also a weak solution in the sense that
it has a locally integrable weak derivative (a
distributional derivative that is a function)
$\phi_{,k}$ satisfying
$$
\int\sqrt{-g}g^{ik}\phi_{,k}\chi_{,i}dV = 0\eqno(1)
$$
for all test functions $\chi$.
  
I then make the following defintions.

$M$ is {\it $L$-globally hyperbolic} if there is a spacelike
surface $S$ such that there is a $1$-$1$ correspondence (defined by taking
the foliation for $\Phi$ to include $S$) between 
Cauchy
data on $S$ (satisfying only local constraints, if any) and global 
solutions
to $L\phi=0$.

A point $p$ in $M$ is {\it $L$-nakedly singular} if it has no
$L$-globally hyperbolic neighbourhood. Otherwise it is called {\it 
$L$-inessential}.

For simplicity of the later exposition I shall take Cauchy data that
is $C^2\times C^1$.
This restriction on the initial conditions is stronger than is really
required: the aim is to illustrate principles, not to obtain the best
possible result.

\bigbreak
{\bf 2. Curve integrable space-times}

Perhaps the most interesting of the singularities where the
differentiability falls below $C^{2-}$ are the shell-crossing
singularities, which are not too unrealistic physically and may be
tractable analytically. With a view to showing that these are
inessential, I shall prove the following:

{\sl{\it Theorem.} Suppose given $(M,g)$ and a point $p$ in $M$ such 
that

{\parindent=40pt
\item{(a)} $g_{ij}$ and $g^{ij}$ are continuous
\item{(b)} $g_{ij}$ is $C^1$ in $M\backslash J^{+}(p)$
\item{(c)} weak derivatives $g_{ij,k}$ exist and are square
integrable on $M$
\item{(d)} the distributionally defined $R^i{}_{jkl}$ is a function
\item{(e)} there is a non-empty open set $C\subset \Real^4$, 
and positive functions\hfil\break $M, N:\Real^{+}\to 
\Real^{+}$ such that, if
$\gamma$ is a curve with $d\gamma/ds \in C$ for all $s$ then
\itemitem{(i)}$\gamma$ is future timelike
\itemitem{(ii)}the integrals 
$$
I_\gamma(a):=\int_0^a\left\vert\Gamma^i_{jk}(\gamma(s))\right\vert
^2
ds\quad \hbox{and}
\quad
J_\gamma(a):= \int^a_0\left\vert R^i{}_{jkl}(\gamma(s))\right\vert ds
$$
\itemitem{}are convergent, with
$$
I_\gamma(a)<M(a),\quad J_\gamma(a)<N(a)\eqno(2)
$$
and $M(a)\to 0$, $N(a)\to 0$ as $a\to 0$.
\par}

Then $p$ is $\dal$-inessential (where $\dal$ is the wave operator).}
\bigskip
The conditions (a) and (c), introduced by Geroch and Traschen [5], are
the minimal conditions for $R^i{}_{jkl}$ to be definable as a
distribution by the usual
coordinate formula in terms of $g_{ij}$ and $g^{ij}$. The set $C$
defines a range of timelike directions which are transverse to any
shocks or caustics that may be present (this is illustrated in the
next section). We refer to (e)
by saying that the space-time is curve-integrable.

I conjecture that the same result holds if the definition of $I$ is
altered to involve the simple modulus of $\Gamma$, rather than the
modulus squared. The proof here, however, requires the stronger
condition above, which essentially asserts that the quadratic and the
linear (in $\Gamma$) parts of the Riemann tensor are separately integrable 
along
the curves considered here. This form of the condition implies, of
course, the square integrability stated separately in (c).

\bigskip
{\it Proof}

\def\t{{\bf t}_{\vec y0}}
\def\kn{\kappa_{\vec y}^{(n)}{}}

\def\tn{{\bf t}^{(n)}_{\vec y}{}}
\def\en#1#2{e_{\vec y}^{(n) #2}{}_{#1}}
\def\enf#1#2{e_{\vec y#1}^{(n) #2}}
\def\Yn{Y_{\vec y\alpha}^{(n)}{}}

We will be able to apply standard theorems on the existence and
uniqueness of solutions to linear partial differential equations,
provided that we can establish an {energy
inequality} for the solution. This in turn will require us to find a
vector field {\bf t} whose covariant derivative is known to be
bounded on spacelike surfaces. We achieve this by taking for the field 
the tangent vector to
a suitable congruence of geodesics. The necessary steps are,
therefore, to show the existence of the congruence and then to compute
the covariant derivative of its tangent vector.

{\sl 2.1 There exists a congruence of timelike geodesics whose tangent
vector has an essentially bounded covariant derivative}

{\it Proof of 2.1}

Since the definition of `inessential' is local, we can shrink $M$ to a
smaller neighbourhood of $p$ if necessary. Since the metric is
continuous we can take a rotation of coordinates so that the surfaces
$S_t=\{x\mid x^0=t\}$ are spacelike in $M$ (shrunk if necessary). 
Suppose
moreover that $p$ is at $t=0$. Choose a fixed vector ${\bf t}_0\in
C$ and let  $S$ be the spacelike surface $S_{t_1}$, where $t_1<0$
is to be determined later. We denote the coordinates $x^\alpha$
($\alpha=1,2,3$) on $S$ by $\vec y$, and let $\t$ denote the vector
with components ${\bf t}_0$ at the point on $S$ with coordinates 
$\vec y$.

Next, we need to establish that the conditions
on the metric are sufficient to ensure the existence of 
geodesics with initial tangents $\t$: we shall take these to form
the required congruence.
Let $S:\Real^4\to \Real$ be a $C^\infty$ smoothing function, i.e.\ 
$S(x)
\geq 0$, $S(x)=S(-x)$, $\int S(x)dx = 1$, $\hbox{support}(S)$ 
compact. Define $S_n(x)=n^4S(nx)$ and let $\Gamma^{(n)i}_{jk} =
S_n*\Gamma^i_{jk}$ (where $*$ denotes convolution). For each 
$\vec y\in
\Real^3$  let $\kn$ be the
$\Gamma^{(n)}$-geodesic with initial tangent vector $\t$ and let
$\tn(s)$ be its tangent vector at parameter $s$.

We note that, for small enough $s$, say $s<s_1$, these tangent vectors 
have components in
$C$ (with $s_1$ being a uniform bound, independent of $n$). This
follows from the geodesic equation
$$
{d\tn^i\over ds} = -\Gamma^{(n)i}_{jk}\tn^j\tn^k\eqno(3)
$$
together with
$$
\left|\int_0^s\Gamma^{(n)i}_{jk}(\lambda(s'))ds'\right|=\left|\int
S_n(z)\left[\int_0^s \Gamma^i_{jk}(\lambda(s')+z)ds'\right]dz\right|\leq 
M(s)^{1/2}s^{1/2}\eqno(4)
$$
for any curve with $\dot\lambda\in C$, which establishes a uniform 
bound for the right hand side of (3).
Indeed, if 
$$
r=\min_{\vec y}\hbox{dist}(\tn(0),C')
$$ 
(where $C'$ is the complement of $C$) and $q = \max_{\vec
y}|\tn(0)|$  then it suffices to
take $s_1$ such that $M(s_1)^{1/2}s^{1/2}<r/(8(q+r)^2)$ which will 
ensure that $|\tn-\t|<r$. 
The time $t_1$ can now be specified as sufficiently close to $0$ to
ensure that $p$ is covered by the curves up to $s_1$.

 We now examine the connecting vector $Y$,
the basic tool being the geodesic deviation
equation for the variation with $\vec y$ of $\kn$. Let $(\en 
{}a)_{a=0\ldots3}$ be a parallely
propagated co-frame on $\kn$ coinciding with the coordinate basis at 
$s=0$, with $\enf b{}$ the corresponding frame, and define
$$
\Yn^a(s):={\partial\kn^i\over\partial y^\alpha}\en ia.
$$
The geodesic deviation equation is then
$$
{d^2\Yn^a\over ds^2} = \en 
iaR^{(n)i}{}_{jkl}\tn^j\tn^k\Yn^{b}\enf bl\eqno(5)
$$
subject to 
$$
\Yn^a(0) = \delta^a_\alpha
$$
and 
$$
{d\Yn^a(0)\over ds} =  \nabla_{\partial_\alpha}\t = 
\Gamma^{(n)a}_{j\alpha}(t_1,\vec y){\bf t}_0^j.
$$

Condition (e) implies that $\en ia$ are uniformly bounded, as are 
$\tn^i$, so that (5) gives
$$
\left| {d^2\Yn^a\over ds^2}\right| \leq Q \sigma \Vert \Yn \Vert,
$$
for some constant $Q$, where
$$
\sigma := \sup_{i,j,k,l}|R^{(n)i}{}_{jkl}|.
$$
It follows that $\Vert Y\Vert$ is bounded by the solution $z$ of the 
majorizing equation
$$
{d^2 z\over ds^2} = K\sigma z\eqno(6)
$$
subject to $z(0) = 1$, $dz(0)/ds = \Vert 
\Gamma^{(n){.}}_{j\alpha}(t_1,\vec y)\Vert =: V$ (defining $V$). 

This equation will imply that $z$, and hence $\Vert Y\Vert$, can be 
bounded in terms of the integral of $R^{(n)}$. Now the significance of 
the conditions (e) is that this integral can be bounded in terms of the 
integral of $R$. Indeed, since the integrals of the linear and quadratic 
parts of $R$ are separately bounded, we have inequalities of the
following form (with $a, b, a_1, a_2$ constants)
$$
\int_0^s|\partial_i\Gamma^{(n)a}_{jk}|ds' \leq \int_0^s|S_n * 
\partial_i\Gamma^{(n)a}_{jk}|ds' \leq \int dz S_n(z) \int_0^s 
|\partial_i\Gamma^{a}_{jk}(\kn(s')+z)|ds'
\leq a(J(s) + I(s))
$$
$$\eqalign{
\int_0^s|\Gamma^{(n)i}_{jk}||\Gamma^{(n)l}_{mp}|ds' &\leq \int dz 
S_n(z) \int dy S_n(y) \int_0^s ds' 
|\Gamma^{i}_{jk}(\kn(s')+z)||\Gamma^l_{mp}(\kn(s')+y)|\cr
&\leq  bI(s)\cr}
$$
and hence
$$
\int_0^s|R^{(n)i}{}_{jkl}|ds' \leq \int_0^s \sigma ds' \leq a_1 M(s) + a_2 N(s) =: 
M_1(s),
$$
say.  To estimate the solution to (6) we then note that if $s_2$ is the 
first value of $s$ at which $|z| = 2$ (possibly $s_2 = \infty$) then 
before $s_2$ 
$$
{d^2 z\over ds^2} \leq 2Q\sigma 
$$
leading, for $0\leq s\leq s_2$, to 
$$
z\leq 1 + Vs + 2Q \int M_1(s) ds.
$$
Thus if we choose $s_0$ so small that
$$
Vs_0 + 2K \int^{s_0} M_1(s) ds < 1
$$
then we will have $z<2$ up to $s_0$, and hence $\Vert \Yn\Vert <2$ 
in this interval.

This bound implies that the function $\kappa: (\vec y, s) \mapsto 
\kn(s)$ is equicontinuous, and so there is by Arzela-Ascoli a 
subsequence that tends to a limit. Choosing this subsequence gives 
meaning to the idea of limiting geodesics.

Having established this, essential boundedness of the derivative of
the tangent vector follows in a similar way.  If  $X^i=X^\alpha\Yn^i+ 
X^0\tn^i$
then 
$$
X^j\tn^i{}_{;j}=X^\alpha\tn^j\Yn^i{}_{;j}=
\Gamma^a_{j\alpha}(\vec y){\bf t}_0^j+\int_0^s\en
iaR^{(n)i}{}_{jkl}\tn^j\tn^k\Yn^lds'
$$
We can now take the limit of this in $L^\infty$ to obtain essential 
boundedness.

{\sl 2.2 Solutions of the wave equation satisfy an energy inequality (12)}

{\it Proof of 2.2}

 The technique closely follows the account of Hawking and
Ellis [6], section 7.4

We are concerned with  solutions (cf (1)) to the wave equation
$$
\dal\phi\equiv g^{ij}\phi_{;ij}=0.\eqno(7)
$$
Suppose initially that $\phi$ is $C^2$ and define 
$$
S^{ij}:= (g^{ik}g^{jl}-{1\over2}g^{ij}g^{kl})\phi_{,k}\phi_{,l}
-{1\over2}g^{ij}\phi^2.
$$
We let $U$ be a compact set bounded to the past by $S$ (i.e.\
$I^{-}(U)\cap\partial U\subset S$) and to the future by a spacelike
surface $H=I^{+}(U)\cap\partial U$.

Working locally, from the continuity of $g^{ij}$ we can choose $U$ to
be foliated by $C^\infty$ 
spacelike surfaces $S_\tau^U$. (We take $0\leq \tau \leq \tau_1$.)

 Set 
$U_\tau=\bigcup_{\tau'<\tau}S_{\tau'}^U$ and define
$$
E(\tau) := \int_{S_\tau^U} S^{ij}t_in_j\sqrt{-g}d^3x
$$
where $n$ is the future normal to $S_t^U$. Our aim is to estimate the
norm 
$$
\Vert\phi\Vert^1_{S,\tau}=\left[\int_{S_\tau^U}(\sum_{i}(\phi_{,i})^2+
\phi^2)d^3x\right]^{1/2}
$$ 
which is related to $E$ by 
$$
kE(\tau)\leq \left(\Vert\phi\Vert^1_{S,t}\right)^2 \leq 
KE(\tau)\eqno(8)
$$
for positive $k, K$. We also introduce 
$$\eqalignno{
\Vert\phi\Vert^1_\tau&=\left[\int_{U_\tau}(\sum_{i}(\phi_{,i})^2+
\phi^2)d^4x\right]^{1/2}\cr
&\leq\left[\int_0^\tau \left( \Vert\phi\Vert^1_{S,t'}\right)^2dt'  
\right]^{1/2}&(9)\cr}
$$ 
and
$$
\Vert\phi\Vert^0_\tau=\left[\int_{U_\tau} \phi^2)d^4x\right]^{1/2}.
$$ 

Stokes' theorem yields
$$
\int_{U_\tau}(S^{ij}t_i)_{;j}\sqrt{-g} d^4x =
(-\int_S+\int_{H_\tau}) S^{ij}t_in_j\sqrt{-g} d^3x.\eqno(10)
$$
By direct calculation
$$
S^{ij}{}_{;j}=(g^{ij}\phi_{,j})(g^{kl}\phi_{;kl}-\phi)
$$
and so the left hand side of (10) becomes (with 
$g^{kl}\phi_{;kl}=\dal\phi$)
$$
\int_{U_t}[(\dal\phi-\phi)\phi_{,k}t^k +
S^{ij}t_{i;j}]\sqrt{-g}d^4x.
$$
Estimating all the terms by the bounds available gives
$$
{1\over K}\left(\Vert\phi\Vert^1_{S,\tau}\right)^2 \leq E(\tau)
\leq E(0) + c\Vert \phi\Vert^1_\tau\Vert L\phi\Vert^0_t
+ c'(\Vert\phi\Vert^1_{t} )^2\eqno(11)
$$
for constants $c, c'$. If $\dal\phi=0$ weakly, this becomes
$$
{1\over K}\left(\Vert\phi\Vert^1_{S,t}\right)^2 \leq E(t)
\leq E(0) 
+ C'(\Vert\phi\Vert^1_{t} )^2.\eqno(12)
$$
while if $E(0)=0$, (11) and (9) give
$$
E(\tau)\leq c_1 (\Vert L\phi\Vert^0_\tau)^2\eqno(13)
$$
for some constant $c_1$.

{\sl 2.3 There exist unique solutions to the wave equation for
$C^2\times C^1$ initial conditions}

{\it Proof of 2.3}

We briefly recall the standard arguments (see, for example [4]) which  
allow us to deduce uniqueness and existence of
 solutions from an energy estimate, using the symmetry of the 
wave operator. 

Uniqueness is immediate: if the difference between two  solutions 
is zero on $S$
then $E(0)=0$ and (13) then implies that the solutions are (pp)
identical. 

Let $V_1$ be the subset of  $L^2(U)$ consisting of $C^\infty$ 
functions that are zero with a zero normal derivative on $H := 
S_{\tau_1}$ and 
let $V_0$ be the subset of  $L^2(U)$ consisting of $C^\infty$ 
functions that are zero with a zero normal derivative on $S := S_{0}$. 
Then the same uniqueness argument holds for both these data conditions, and we 
have equation (13), which implies there exists a constant $c_2$ such that
$$
\Vert \phi \Vert \leq C_2 \Vert L\phi \Vert\eqno(14)
$$
for $\phi \in V_1$ (from now on all norms are in $L^2$).

To prove existence subject to conditions $\phi=\phi_0, 
\dot\phi=\phi_1$ on $S$, choose a $C^2$ function $f$ satisfying these 
conditions and look for a function $\psi = \phi - f$ satisfying zero 
boundary conditions and $L\psi = -Lf =: \chi$. The required fucntion 
$\psi$ will satisfy
$$
\int_U\psi Lw d^4x = \int_U \chi w d^4x
$$
for all $w\in V_1$. From (14) 
$$
\left|\int_U\chi w d^4x\right | \leq c_2 \Vert\chi\Vert\Vert Lw\Vert
$$
so that the map $k:Lw \mapsto \int_U\chi wd^4x$ is a bounded linear 
functional on $LV_1$. But $LV_1$ is dense in $L^2$, because $V_0$ 
is dense and if $\int_U \phi Lw d^4x =0$ for all $Lw\in LV_1$ and 
$\phi\in V_0$ then we must have $\phi=0$. So $k$ defines an element 
$\psi$ of $L^2$ such that $\int_U\psi Lwd^4x = k(Lw) = \int_U\chi 
wd^4x$; so that $\phi=f+\psi$ is the required  solution.

This concludes the proof.
\bigskip
{\bf 3. Application to dust caustic (shell crossing) space-times}

Though there is as yet no rigorous proof,  there are very strong 
indications [3] that shell-crossing spherically symmetric dust 
configurations produce relativistic solutions in which the flow lines of 
matter produce a caustic, as in the gravity-free case (see figure 1). 

\midinsert
\centerline{\psfig{file=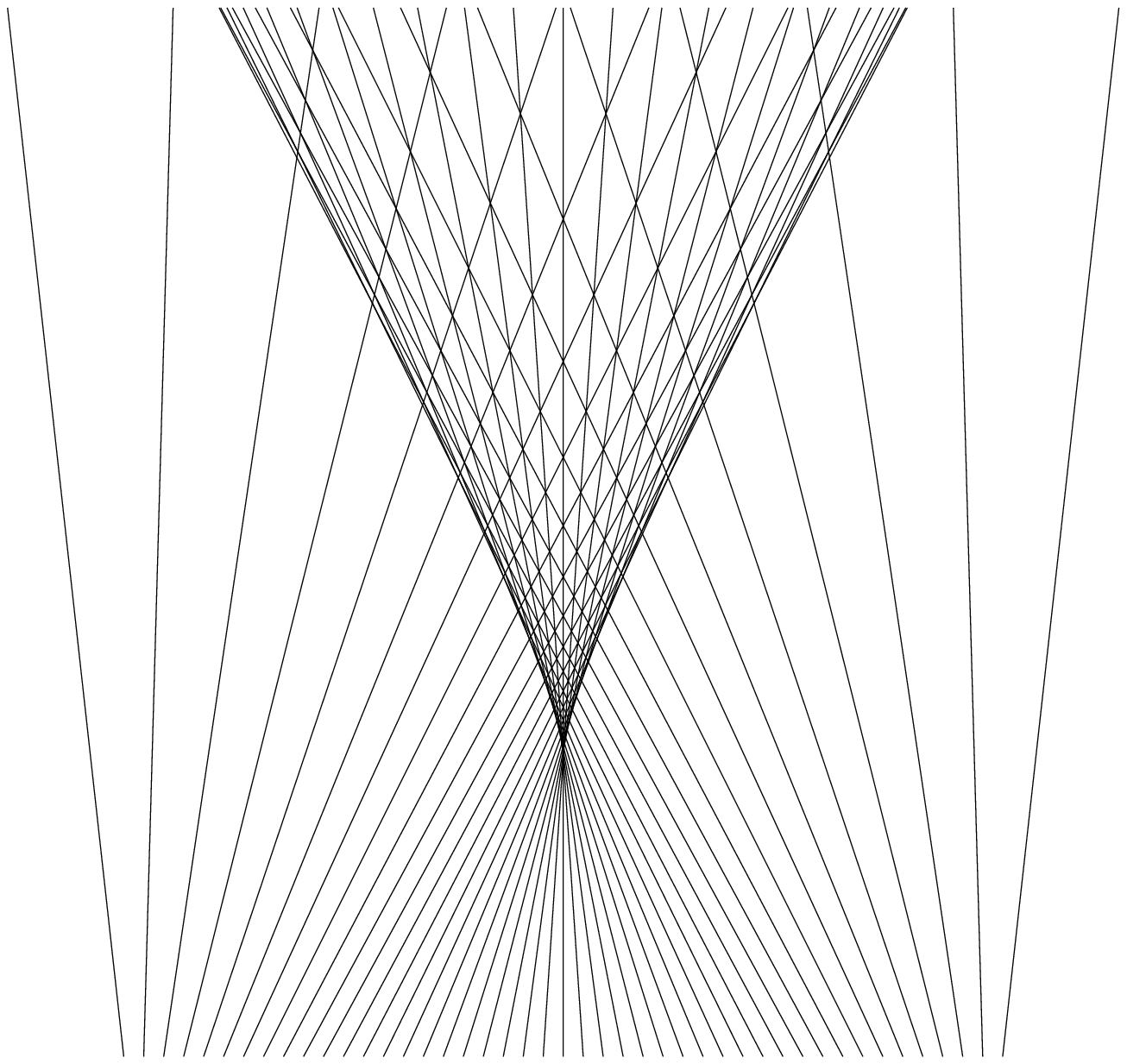,height=7cm}}
\centerline{Figure 1: space-time diagram of world lines of shell-crossing dust}
\endinsert

In 
what follows I shall be assuming the existence of such a space-time, in 
which the general form of the matter density is the same as that in the 
gravity free case. Since the matter density determines the Riemann 
tensor and the connection via simple integrals in this case, we can pass 
from the density to the Riemann tensor immediately.

From catastrophe theory, the generic gravity free caustic is 
diffeomorphic to the following canonical form. If $(r,t)$ are the 
essential coordinates in a spherically symmetric situation, and the flow 
lines of matter are parametrised by $t$, with $v:=dr/dt$ constant on 
each line, then the lines when lifted to curves $t\mapsto (r,t,v)$ in 
$\Real^3$ (thought of as a reduction of the tangent bundle) rule the surface 
$\Sigma$ with equation $r = f(v,t) := vt -av^3$ for a constant $a$. The 
projection in the tangent bundle corresponds to the projection 
$p:(r,t,v)\mapsto (r,t)$,  and the critical points of $p|\Sigma$
constitute the 
curve $t = 3av^2$ in $\Sigma$. The caustic is the image under $p$ of this critical 
point set, namely $t=3a^{1/3}(r/2)^{2/3}$. Each point on the ruled 
surface makes a contribution $\rho = \rho_0(v)(\partial f/\partial v)^{-
1}= \rho_0/(t- 3av^2)$ to the total density at the corresponding point 
of space-time, for some function $\rho_0(v)$ giving the density 
distribution in velocity-space.

Integrability of the Riemann tensor along curves will depend on its 
behaviour near the caustic. Consider, therefore, a coordinate straight 
line cutting the caustic at the image of a point on the critical point
set in $\Sigma$ with velocity 
$v_0$, i.e. at $r_0= 2av_0^3$, $t_0=3av_0^2$, the line being $x = 
x_0+\lambda(t-t_0)$. Setting $t=t_0+\tau$, $v=v_0+\nu$ and working 
to lowest significant order in $\tau$ and $\nu$, we obtain
$$
\nu \approx \left({v_0-\lambda\over3av_0}\right)^{1/2}\tau^{1/2}
$$
for $v_0\neq0$ (where the condition $v_0>\lambda$ is required for 
transversality to the caustic) and
$$
\nu \approx -\left({\lambda\tau\over a}\right)^{1/3}
$$
for $v_0=0$. The key point arising from this as a consequence is 
that $\rho$ is 
integrable along the curve, a result which is diffeomorphism invariant 
and so applies to the generic caustic.

Passing to the relativistic case, as previously noted we assume that this 
behaviour of the density still holds, specifically when the metric is 
presented in coordinates linearly related to double null coordinates. 
(As described in detail in [3], the choice of coordinates becomes 
significant in general relativity, as opposed to the Newtonian case, 
because coordinate transformations  -- for instance, from curvature 
coordinates to double null coordinates -- are typically specified by 
geometrical conditions and so are not, in this case, $C^\infty$.) With 
this assumption, the Riemann tensor is curve-integrable and (see the 
treatment of these coordinates in [1]) the connection coefficients are 
bounded. It would then be the case that the caustic is not a 
$\dal$-essential singularity, so that cosmic censorship is not violated.

{\bf References}

[1] Clarke C.J.S. {\it The Analysis of Space-Time Singularities},
Cambridge University Press, 1993 

[1] Clarke, C J S ``Singularities: boundaries or internal points'', {\it
Proceedings of the International Conference on Gravitation and
Cosmology 1995}

[3] Clarke C.J.S. and O'Donnell N. ``Dynamical extension through a
space-time singularity,'' {\it Rendiconti del seminario matematico,
Universit\'a e Politecnico Torino} {\bf 50}, (1)  39--60, 1992

[4] Egorov, Yu V and Shubin, M A {\it Partial Differential Equations
I}, Springer Verlag, 1992

[5] Geroch, R and Traschen, J. {\it Phys. Rev.}
D{\bf36}, 1017-31, 1987.

[6] Hawking, S W and Ellis, G F R {\it The large scale structure of
space-time}, Cambridge University Press, 1973

[7] Penrose R. ``Singularities and Time Asymmetry,'' in {\it General
Relativity. An Einstein Centenary Survey.} ed S.W. Hawking and W.
Israel, Cambridge University Press, 1979

\end